\documentclass[sigconf]{acmart}

\AtBeginDocument{
  }

\setcopyright{acmlicensed}
\copyrightyear{2026}
\acmYear{2026}
\acmDOI{XXXXXXX.XXXXXXX}
\acmConference[ACM The Web Conference '27]{Make sure to enter the correct
  conference title from your rights confirmation email}{May 10--14,
  2027}{Dublin, Ireland}
\acmISBN{978-1-4503-XXXX-X/2018/06}

\usepackage{algorithmic}
\usepackage{graphicx}
\usepackage{textcomp}
\usepackage{xcolor}
\usepackage{array}
\usepackage{soul}
\usepackage{tabularx,multirow}

\usepackage[most]{tcolorbox}
\usepackage{placeins}
\newcommand{\finding}[2]{
\begin{center}
\begin{tcolorbox}[
    width=.9\linewidth,
    colback=white,
    colframe=black,
    arc=1mm,
    leftrule=2pt,
    rightrule=0.5pt,
    toprule=0.5pt,
    bottomrule=0.5pt,
    left=5pt,
    right=5pt,
    top=5pt,
    bottom=5pt
]
\textbf{Finding #1.} \textit{#2}
\end{tcolorbox}
\end{center}
}

\begin{document}

\title{Domain Decoupling Attack: Exploiting the Validation Gap Between Protective DNS and Shared Edge Routing}

\author{Weizhe Wang}
\affiliation{
  \institution{Tianjin University}
  \city{Tianjin}
  \country{China}}
\author{Minghong Dong}
\affiliation{
  \institution{Tianjin University}
  \city{Tianjin}
  \country{China}}
\author{Jinhao Li}
\affiliation{
  \institution{Tianjin University}
  \city{Tianjin}
  \country{China}}
\author{Lyuchuan Wei}
\affiliation{
  \institution{Tianjin University}
  \city{Tianjin}
  \country{China}}
\author{Yao Zhang}
\affiliation{
  \institution{Tianjin University}
  \city{Tianjin}
  \country{China}}
\author{Qiang Hu}
\affiliation{
  \institution{Tianjin University}
  \city{Tianjin}
  \country{China}}
\author{Qi Guo}
\affiliation{
  \institution{Tianjin University}
  \city{Tianjin}
  \country{China}}
\author{Guangquan Xu}
\affiliation{
  \institution{Tianjin University}
  \city{Tianjin}
  \country{China}}
\author{Bin Wu}
\affiliation{
  \institution{Tianjin University}
  \city{Tianjin}
  \country{China}}

\renewcommand{\shortauthors}{Wang et al.}

\begin{abstract}
Command-and-control, data transfer, and censorship circumvention can hide in traffic to widely used services. DNS-bound access control grants permits based on resolved IP addresses, while shared edges route requests by tenant identity, creating a validation gap. An attacker can resolve an allowed anchor, obtain a permit for its shared edge address, and access a hidden tenant by presenting the tenant's name as both SNI and HTTP Host. To investigate this gap, we present the Domain Decoupling Attack (DDA), which exploits this gap in CDN and non-CDN hosting while preserving SNI--Host consistency. Across 1,069,048 domains, we conduct 18,025,068 successful probes. Overall, 95.8\% of domains accept an external SNI, and 84.6\% return content matching the named tenant. Exposure reaches 99.26\% across 506,763 CDN domains and 92.75\% across 562,285 non-CDN domains. The co-location graph places 666,413 exposed domains in cross-tenant components; 97.7\% of 54,069 cross-tenant non-CDN addresses have a confirmed path to another tenant. None of the tested defenses bind tenant identity to the DNS-approved name. Cleartext inline IPS can block access based on SNI, but Encrypted Client Hello (ECH) hides it. The Windows Zero Trust DNS (ZTDNS) permit remains reusable in both cases, enabling the concealment of command-and-control and data-exfiltration traffic and facilitating censorship circumvention. The risk conceals command-and-control and data-exfiltration traffic and supports censorship.
\end{abstract}



\received{20 February 2007}
\received[revised]{12 March 2009}
\received[accepted]{5 June 2009}

\maketitle

\section{Introduction}

Network attackers conceal command-and-control communication, data transfer, and censorship-circumvention traffic within traffic to widely used Internet services~\cite{zhang2016hunting,zorawski2023long,anderson2023assessing,fifield2015blocking,sommer2019deniable,yao2023hiding}. Content delivery networks (CDNs) are well suited to this purpose: their edge servers serve many independent domains and carry substantial legitimate traffic, making address-level blocking disruptive to benign services~\cite{wei2021domain,wails2025censorship}. Existing CDN-based evasion techniques exploit specific provider behaviors. Domain fronting separates the TLS server name indication (SNI) from the HTTP Host~\cite{fifield2015blocking,xie2025fakeapp} and has been studied as a covert channel~\cite{xie2024domeye}. Providers have responded by introducing SNI–Host consistency checks~\cite{lin2024detecting,zheng2024reqsminer}. Domain borrowing exploits gaps in domain-ownership verification during onboarding~\cite{cimaszewski2023effective,ding2021domain}, while domain shadowing relies on provider-internal front-to-backend rewrites~\cite{ding2021domain,wei2021domain}. Although providers vary in how they restrict these techniques~\cite{subramani2024discovering}, each technique depends on behavior that a provider can restrict.

DNS-bound access control introduces an additional authorization decision. Protective DNS determines which domain names a client may resolve~\cite{rose2020zero}. Windows ZTDNS blocks outbound IP traffic by default, permitting traffic to addresses returned by a trusted DNS service~\cite{microsoft2025ztdns}. In effect, this design translates approval of a domain name into permission to connect to an IP address. The resulting permission is only as specific as the name-to-address mapping. Shared addressing is a standard deployment model. One address serves many domains through name-based virtual hosting and CDN edges~\cite{rfc7230,shue2007web,cloudflare-ip}, and the mapping between names, addresses, and services is many-to-many~\cite{scott2016satellite,huitema2020issues}. A shared edge selects a tenant using the server name presented in the TLS SNI or, under ECH, the inner SNI~\cite{Rescorla2026ECH}. We refer to this name as the connection identity. The DNS policy evaluates the resolved name, while the edge routes the connection using the connection identity. The address is the only value shared by the two decisions.

\begin{figure}
    \centering
    \includegraphics[width=\linewidth]{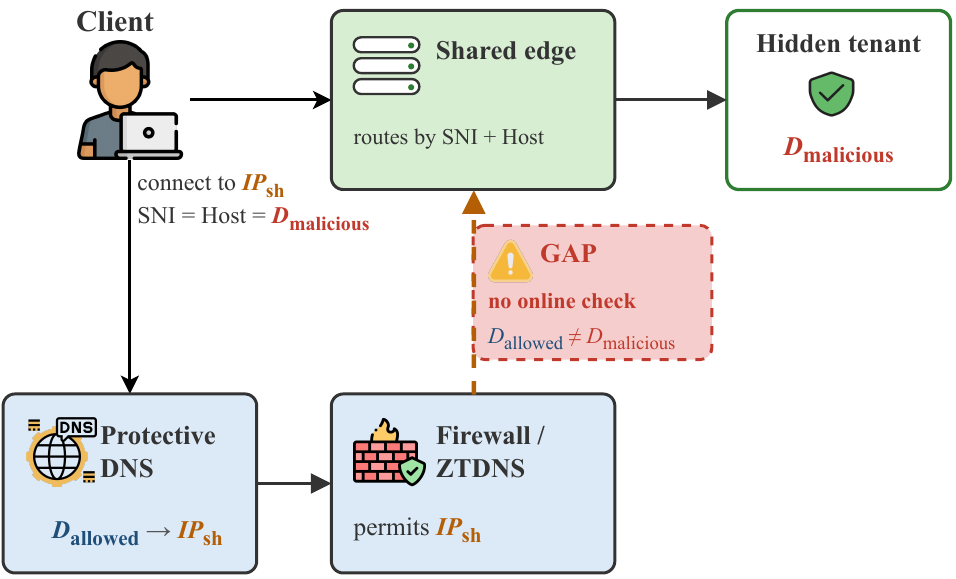}
    \caption{Validation gap between DNS-based authorization and shared-edge tenant routing.}
    \label{fig:motivation}
\end{figure}

We refer to this separation as a validation gap.. Consider an allowed domain \(D_{\mathit{allowed}}\) that resolves to a shared edge address \(IP_{\mathit{sh}}\), and a hidden tenant \(D_{\mathit{malicious}}\) served at the same address. The client resolves the anchor \(D_{\mathit{allowed}}\), and \(IP_{\mathit{sh}}\) enters the endpoint's permitted address set. It then connects to that address and presents \(D_{\mathit{malicious}}\) as both the connection identity and the HTTP Host. The shared edge routes the connection to \(D_{\mathit{malicious}}\). But no online check in this decision chain verifies whether the connection identity is authorized under the DNS policy that granted access to the address. Figure~\ref{fig:motivation} shows these two decisions and the address that links them.

We call this attack the Domain Decoupling Attack (DDA). Because the connection identity matches the HTTP Host, DDA passes SNI--Host consistency checks. It requires neither onboarding a high-reputation domain the attacker does not own nor a provider-internal rewrite. Instead, DDA uses different names for DNS resolution and connection establishment. The gap arises when otherwise correct decisions are combined: the DNS policy approves an allowed name, the endpoint permits the resulting address, and the edge routes the connection to the requested tenant. What is missing is a check that binds the connection identity to the DNS authorization that granted access to the address. That check requires information from both decisions, but neither decision point enforces it. Correlating DNS and connection logs can associate the name that authorized access to the address with the name presented on the subsequent connection after the fact; it does not enforce the binding when the connection is admitted. DDA requires only an ordinary DNS lookup followed by an HTTPS connection.

This gap is not specific to CDNs. It can arise wherever DNS-bound authorization is combined with an address that serves multiple domains and selects a tenant at connection time. Non-CDN shared hosting, reverse proxies, and multi-tenant load balancers follow the same pattern; CDN edges are its largest and most visible example. We therefore measure both CDN and non-CDN deployments. Starting from the Tranco Top 1M list~\cite{le2019tranco}, our measurement covers 1,069,048 domains and yields 18,025,068 successful probe records. Of these domains, 1,024,540 (95.8\%) accept a TLS connection with an external SNI, and 904,789 (84.6\%) return content matching the tenant named in the connection. Exposure reaches 99.26\% among the 506,763 CDN domains and 92.75\% among the 562,285 non-CDN domains. In a graph that connects domains sharing an address, 666,413 exposed domains belong to components spanning multiple registrable domains. Of the 54,069 non-CDN addresses serving multiple registrable domains, 52,811 (97.7\%) provide at least one confirmed path to another tenant. These findings show that the gap extends beyond CDNs to shared hosting more broadly.

The tested defenses address parts of this path, but none closes the gap. Zeek and Suricata inspect the visible connection identity, OPNsense and pfSense enforce address-level policy, and Windows ZTDNS converts a trusted DNS decision into an address permit. Under a cleartext SNI, Suricata in inline IPS mode can drop the hidden tenant because it sees the name. This is the only tested configuration that blocks it selectively. Under ECH, the inner name is encrypted, and a path observer retains the outer name, the shared address, and the ECH extension. ZTDNS blocks a direct connection to the hidden tenant, but resolving the anchor places the shared address in the permitted set. An ordinary TLS connection reaches the hidden tenant through that permit. A Windows-native ECH client reaches it as well, while path observers see only the outer name. The inspection tools lack the DNS-approved name, and the authorization controls lack the tenant identity. A domain-level decision becomes an address-level permit, and the permit survives a change of tenant.

The operational consequences follow directly. A compromised endpoint can send command-and-control and data-exfiltration traffic to an address that an allowed anchor has already placed in the permitted set. The same separation supports censorship circumvention. A client in a censored network can resolve an allowed anchor on shared infrastructure and then reach a blocked tenant at the same address. Against DNS-based filtering, the censor sees only the anchor at the DNS layer, and ECH extends that separation to on-path observers that inspect the connection identity.

This paper makes the following contributions.

\begin{itemize}

\item We present Domain Decoupling Attack (DDA), an attack that exploits the validation gap between DNS-based authorization and shared-edge tenant routing. DDA obtains an address permit through an allowed anchor and reaches a different tenant at the same address while keeping the connection identity and HTTP Host consistent. It avoids the SNI--Host mismatch of domain fronting, the ownership-verification gap of domain borrowing, and the provider rewrite of domain shadowing.

\item We measure the exposure of the gap across 1,069,048 domains and 18,025,068 successful probe records. Exposure reaches 95.8\% overall, and 84.6\% of domains reach the HIGH level, where the edge returns content matching the tenant named in the connection. The rate is 99.26\% for CDN and 92.75\% for non-CDN domains. The non-CDN result extends the gap beyond CDN hosting.

\item We identify trust propagation across shared addresses, where a permit for one shared address also admits connections to other tenants at that address, and shared addresses link these tenants into larger components. The same-address graph contains 666,413 exposed domains in components that span more than one registrable domain, and 97.7\% of 54,069 non-CDN addresses that serve multiple registrable domains contain at least one confirmed path to another tenant.

\item We evaluate representative network defenses and show that none of the tested controls binds the connection identity to the DNS-approved name. Cleartext inline IPS can block the hidden tenant, ECH removes that check, and the ZTDNS address permit remains reusable in both visibility conditions. The result identifies the cross-plane binding check that DNS-bound access control requires on shared addresses.

\end{itemize}

\section{Domain Decoupling Attack}

\subsection{System and Threat Model}\label{sec:model}

\begin{figure*}
    \centering
    \includegraphics[width=0.9\linewidth]{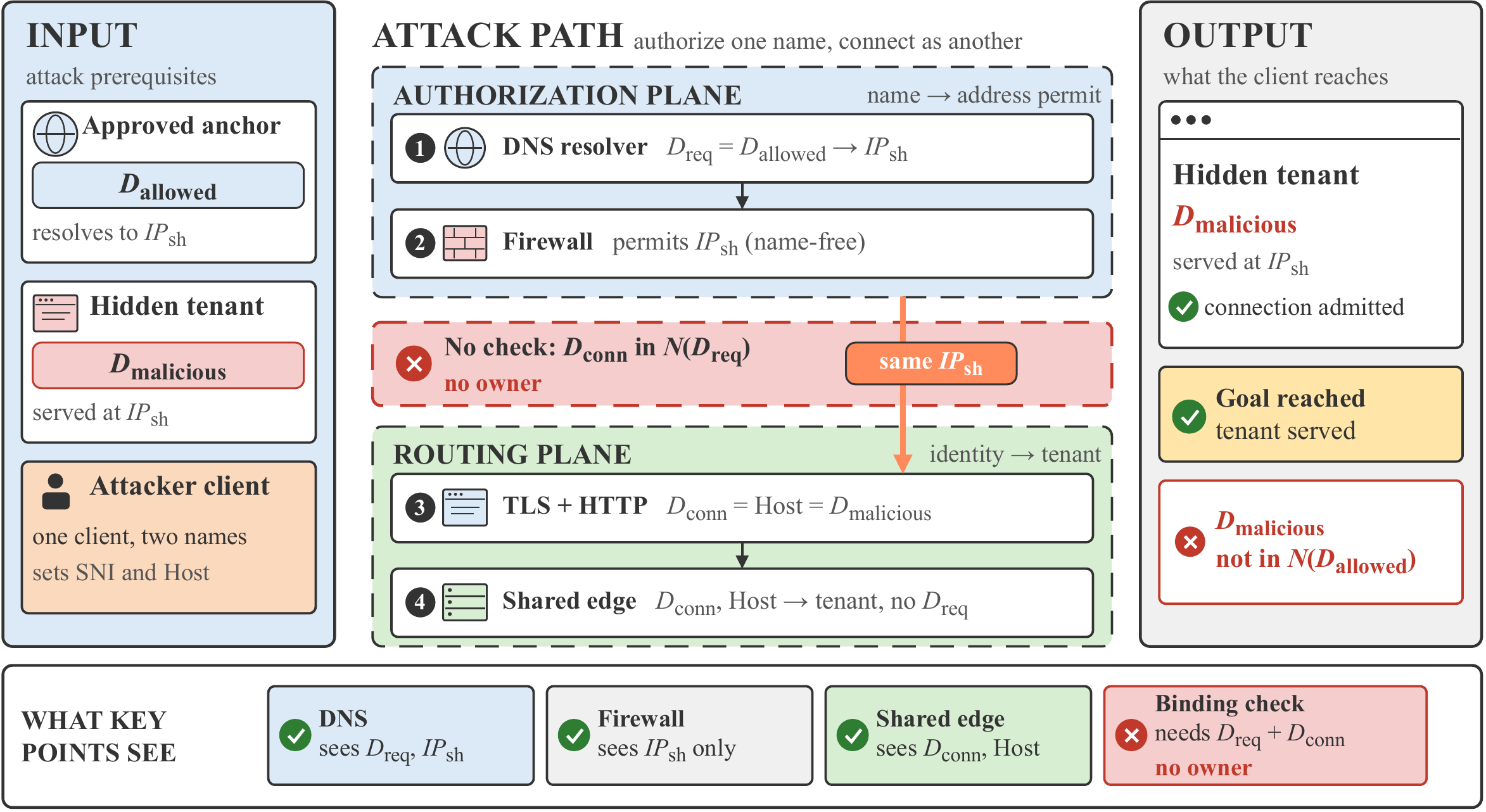}
    \caption{Overview of DDA. The authorization plane approves \(D_{\mathit{allowed}}\) and yields a name-free address permit for \(IP_{\mathit{sh}}\). The routing plane selects a tenant from \(D_{\mathit{conn}}\) and \(Host\). The planes meet at \(IP_{\mathit{sh}}\). The binding predicate \(D_{\mathit{conn}}\in N(D_{\mathit{req}})\) has no owner.}
    \label{fig:overview}
\end{figure*}

An address is \emph{shared} if it serves more than one domain. We write \(IP_{\mathit{sh}}\) for a shared address used in a connection. A CDN edge is the most visible example, but name-based virtual hosting, reverse proxies, and multi-tenant load balancers have the same property. A communication attempt to a shared address is
\begin{equation}
L=(D_{\mathit{req}},IP_{\mathit{sh}},D_{\mathit{conn}},Host),
\end{equation}
where \(D_{\mathit{req}}\) is the name submitted for resolution (\(\bot\) if none), \(D_{\mathit{conn}}\) is the identity the edge uses to select a tenant, and \(Host\) is the HTTP Host value. Without ECH, \(D_{\mathit{conn}}\) is the cleartext SNI. With ECH, it is the inner SNI recovered by the edge, and an on-path observer sees only the outer public name.

Three controls decide whether \(L\) is admitted. A DNS policy \(P_{\mathit{dns}}\) evaluates \(D_{\mathit{req}}\). An address policy \(P_{\mathit{fw}}\), enforced by a firewall, evaluates \(IP_{\mathit{sh}}\). Where deployed, an inspection policy \(P_{\mathit{dpi}}\) evaluates the identity visible to it. A connection is admitted when all deployed policies allow it and the shared edge routes it to a served tenant.

We group these decisions into two functional planes. The authorization plane (\(P_{\mathit{dns}}\) and \(P_{\mathit{fw}}\)) turns an approved name into a permit for an address. The routing plane comprises the visible connection identity, \(P_{\mathit{dpi}}\) where deployed, and the shared edge, and it turns a connection identity into a tenant. The planes are analytical groupings, not protocol layers. At the decision boundary, the only value they share is \(IP_{\mathit{sh}}\). The name that enters the authorization plane does not reach the routing plane, and the identity that selects the tenant does not return to the authorization plane.

A DNS-based control should bind the connection to the domain that justified the address permit. For a connection that starts with DNS resolution, we call the deployment \emph{domain-binding} if every admitted connection satisfies
\begin{equation}\label{eq:binding}
D_{\mathit{conn}}=Host
\quad\text{and}\quad
D_{\mathit{conn}}\in N(D_{\mathit{req}}),
\end{equation}

where \(N(D_{\mathit{req}})\) is the set of names authorized by an approval of \(D_{\mathit{req}}\). Normally \(N(D_{\mathit{req}})=\{D_{\mathit{req}}\}\), and it may include configured aliases. A \emph{validation gap} exists when an admitted connection satisfies \(D_{\mathit{conn}}=Host\) but violates \(D_{\mathit{conn}}\in N(D_{\mathit{req}})\).
The first component is a consistency check internal to the routing plane, where any point that observes both \(D_{\mathit{conn}}\) and \(Host\) can check it. The second is the cross-plane join. It requires \(D_{\mathit{req}}\) from the authorization plane and \(D_{\mathit{conn}}\) from the routing plane, and no single decision point holds both. The validation gap is this missing cross-plane check.

\textbf{Attacker.} The attacker controls a client application and a destination domain \(D_{\mathit{malicious}}\) served by a shared edge. It chooses an allowed domain \(D_{\mathit{allowed}}\) that resolves to the edge address serving \(D_{\mathit{malicious}}\). We call \(D_{\mathit{allowed}}\) the \emph{anchor} and \(D_{\mathit{malicious}}\) the \emph{hidden tenant}. Controlling the anchor is not necessary. The anchor only needs to be an approved name that resolves to the same edge and is not authorized for the hidden tenant. The attacker can query the endpoint's trusted resolver and controls \(D_{\mathit{conn}}\) (including ECH) and \(Host\). It cannot modify any deployed policy, alter the resolver's response, obtain the edge private key, or register or modify \(D_{\mathit{allowed}}\) in the provider's control plane.

\textbf{Conditions.} The construction relies on four conditions.
\begin{itemize}
  \item[\textbf{A1.}] (Shared address) \(D_{\mathit{allowed}}\) resolves to \(IP_{\mathit{sh}}\), and the edge also serves \(D_{\mathit{malicious}}\) at that address.
  \item[\textbf{A2.}] (Address-level permit) Approving \(D_{\mathit{allowed}}\) places \(IP_{\mathit{sh}}\) in the firewall's permitted address set. The permit is not tied to the name used by a later connection.
  \item[\textbf{A3.}] (Connection-identity routing) The edge selects the tenant from \(D_{\mathit{conn}}\) and \(Host\), and does not consult \(D_{\mathit{req}}\).

  \item[\textbf{A4.}] (Controllable, consistent identity) The client can set \(D_{\mathit{conn}}=Host\) to the hidden tenant's name \(D_{\mathit{malicious}}\). Under ECH, \(D_{\mathit{conn}}\) is the inner SNI, and the outer name is the public ECH name.
\end{itemize}

A1 defines the shared interface. A2 is the authorization-plane condition, and A3 and A4 are the routing-plane and client-side conditions.
None of the four conditions is specific to CDN infrastructure, and any shared endpoint that routes by connection identity can satisfy them.

\subsection{Attack Overview}\label{sec:overview}

Figure~\ref{fig:overview} shows the two planes and the only value they share at the decision boundary. The canonical DDA instance is
\begin{equation}
L^{\star}=(D_{\mathit{allowed}},IP_{\mathit{sh}},
D_{\mathit{malicious}},D_{\mathit{malicious}}),
\end{equation}
where \(D_{\mathit{malicious}}\notin N(D_{\mathit{allowed}})\). The client acts as a legitimate client of \(D_{\mathit{allowed}}\) in the authorization plane and as a legitimate tenant of \(D_{\mathit{malicious}}\) in the routing plane. The two roles meet only at \(IP_{\mathit{sh}}\). No decision point in either plane observes both.

The planes answer different questions. The authorization plane decides whether the client may reach \(IP_{\mathit{sh}}\), and the routing plane decides which tenant the connection belongs to. Neither decides whether the connection is authorized for that tenant. That is the domain-binding check, and no decision point owns it. Each local decision is correct. The gap appears only when the decisions are composed, and this is the validation gap that DDA exploits.

Because the client keeps \(D_{\mathit{conn}}=Host\), TLS and HTTP agree, and the connection is internally consistent. The decoupling lies between the name that authorized the address and the identity that selects the tenant. Unlike domain fronting, which exposes an SNI--Host mismatch, DDA places the mismatch across the two planes. The permit is address-level and carries no tenant identity, so the trust placed in \(D_{\mathit{allowed}}\) propagates to every other tenant served at \(IP_{\mathit{sh}}\).

\subsection{Domain Decoupling Attack}\label{sec:procedure}

\begin{table*}[ht]
\centering
\caption{DDA across the authorization and routing planes. Each point makes a locally correct decision. The cross-plane binding has no owner.}
\label{tab:planes}
\small
\renewcommand{\arraystretch}{1.25}
\setlength{\tabcolsep}{8pt}
\begin{tabularx}{0.8\textwidth}{
    @{}
    >{\raggedright\arraybackslash}p{0.1\textwidth}
    >{\raggedright\arraybackslash}p{0.1\textwidth}
    >{\raggedright\arraybackslash}X
    >{\raggedright\arraybackslash}X
    @{}
}
\toprule
Plane & Point & What it sees & What it cannot check \\
\midrule
\multirow{2}{*}{\textit{Authorization}}
& DNS
& \(D_{\mathit{req}},\,IP_{\mathit{sh}}\)
& \(D_{\mathit{conn}}\) \\
& Firewall
& \(IP_{\mathit{sh}}\)
& \(D_{\mathit{req}},\,D_{\mathit{conn}}\) \\
\midrule
\multirow{2}{*}{\textit{Routing}}
& Path DPI
& The visible connection name
& \(D_{\mathit{req}}\) and, under ECH, the inner SNI \\
& Shared edge
& \(D_{\mathit{conn}},\,Host\)
& \(D_{\mathit{req}}\) \\
\midrule
\multicolumn{2}{@{}l}{\textbf{Binding check}}
& \(D_{\mathit{req}}\) at DNS and \(D_{\mathit{conn}}\) at routing
& \(D_{\mathit{conn}}\in N(D_{\mathit{req}})\): no owner \\
\bottomrule
\end{tabularx}
\end{table*}

DDA is a construction that spans both planes. It obtains an address permit under one name, then selects a tenant at that address under another. The authorization plane sees the anchor, and the routing plane sees the hidden tenant, so neither sees both. The connection is internally consistent (\(D_{\mathit{conn}}=Host\)) and violates the cross-plane component of Eq.~\ref{eq:binding}.

DDA has two halves, one per plane. In the authorization half, the attacker selects an anchor \(D_{\mathit{allowed}}\) that \(P_{\mathit{dns}}\) approves and that resolves to \(IP_{\mathit{sh}}\). The hidden tenant \(D_{\mathit{malicious}}\) is served at \(IP_{\mathit{sh}}\), and \(D_{\mathit{malicious}}\notin N(D_{\mathit{allowed}})\). The client resolves the anchor, and the resolver applies \(P_{\mathit{dns}}\), approves the name, and returns \(IP_{\mathit{sh}}\). \(P_{\mathit{fw}}\) then places \(IP_{\mathit{sh}}\) in its permitted set.
In the routing half, the client opens a TCP connection to \(IP_{\mathit{sh}}\). It presents \(D_{\mathit{malicious}}\) as \(D_{\mathit{conn}}\) in the TLS ClientHello and as \(Host\) in the HTTP request. The shared edge terminates TLS, selects the tenant from these values, and forwards the request to the \(D_{\mathit{malicious}}\) backend. Under ECH, \(D_{\mathit{conn}}\) is the inner SNI, and the outer name is not the connection identity.

The address permit is checked and applied when the SYN arrives, and a single connection does not exhaust it. Within its lifetime, later connections reuse the same address-level decision, whatever name they present in TLS. The permit records neither the anchor nor the tenant. The client can therefore repeat the construction or open follow-on connections without repeating the DNS step.

Table~\ref{tab:planes} shows what each decision point sees in the same construction. It traces one flow through both planes.

\textbf{Why every deployed check passes.} Every decision in Table~\ref{tab:planes} is correct for the question it answers. The authorization plane answers whether the client may reach \(IP_{\mathit{sh}}\), but it never sees \(D_{\mathit{conn}}\). The routing plane answers which tenant the connection belongs to, but it never sees \(D_{\mathit{req}}\). The binding predicate \(D_{\mathit{conn}}\in N(D_{\mathit{req}})\) is the only question that requires both, and no decision point owns it. \(P_{\mathit{fw}}\) admits an address, while the edge selects a tenant, so neither decision is the binding check.

The first component of Eq.~\ref{eq:binding} is checkable where both names are visible. The client keeps \(D_{\mathit{conn}}=Host\), so any SNI--Host consistency check passes. The second component is the cross-plane join. At the resolver, \(D_{\mathit{req}}\) is known but \(D_{\mathit{conn}}\) is not, and the firewall sees only \(IP_{\mathit{sh}}\). An on-path inspector sees the visible connection name but not \(D_{\mathit{req}}\). Under ECH, it sees only the outer name. The edge sees \(D_{\mathit{conn}}\) and \(Host\) but never receives \(D_{\mathit{req}}\). Log correlation can reconstruct the pair afterwards, but admission is online and distributed. What remains is an information partition across enforcement points, not a missing feature in one product.

Under A1--A4, the construction yields a connection with \(D_{\mathit{conn}}=Host=D_{\mathit{malicious}}\) and \(D_{\mathit{malicious}}\notin N(D_{\mathit{allowed}})\). If no deployed policy independently blocks \(D_{\mathit{malicious}}\), every policy admits it on the evidence it holds, and Eq.~\ref{eq:binding} is violated. An independent blocklist can reject the hidden tenant, but that rejection is not the cross-plane check. In this sense, DDA exposes a compositional gap. Each plane is locally sound, but their composition is not.

\textbf{Scope and visibility.} The construction is fixed, but deployment properties change what an observer can see and how often the authorization half must be repeated. ECH changes what is visible on the routing path. When the edge publishes an ECHConfig for \(D_{\mathit{malicious}}\), the outer ClientHello carries the ECHConfig's public name, and the inner ClientHello carries \(D_{\mathit{malicious}}\). The edge decrypts the inner SNI and still selects the tenant from \(D_{\mathit{conn}}\), while a path observer without the edge key sees only the outer name. That name is not \(D_{\mathit{conn}}\), so ECH changes who can observe the connection identity, not what the edge knows. Because the edge still lacks \(D_{\mathit{req}}\), the cross-plane predicate remains unowned.

Permit and state lifetimes change repetition and inspection, not the construction. The address permit is stored per address and checked at the SYN, and it is not consumed by a single connection. Later connections can therefore reuse it without repeating the DNS step. A stateful inspector may retain state for an address after a first allowed connection, and a later connection may then receive a reduced or reused decision, which can avoid an independent inspection of that flow. Neither property creates a new attack. The construction is the same, and the cross-plane predicate is still unowned.

The scope boundary is the absence of the authorization plane. If clients are not required to use trusted DNS, \(D_{\mathit{req}}=\bot\) and Eq.~\ref{eq:binding} does not apply. The same address-and-tenant decoupling may still be possible, but it is not DDA. DDA is defined by the composition of an authorized address with a different connection identity. We therefore treat direct IP as a scope boundary, not as a variant of the construction.

The construction is unavailable when no approved anchor is co-located with the hidden tenant. The connection is rejected when a deployment binds the connection identity to the authorization name, rejects an external connection identity, or independently blocks the hidden tenant. These are the properties a deployment must enforce, and the conditions our evaluation probes.

Within the trusted-DNS scope, the invariant does not change. The two planes meet at \(IP_{\mathit{sh}}\). The name that authorized the address never reaches the routing decision, and no online decision point owns \(D_{\mathit{conn}}\in N(D_{\mathit{req}})\). ECH changes visibility, and state can prevent an inspection from repeating. Direct IP removes the authorization plane, but none of these changes the missing join.

\section{Evaluation}\label{sec:evaluation}

\subsection{Evaluation Setup}

\textbf{Measurement dataset.} We derive the evaluation dataset from the Tranco Top 1M list~\cite{le2019tranco}. A breadth scan covers all listed apex domains, and a depth scan covers the top 3,000 domains and their collected subdomains. After deduplication and removal of unreachable domains, 1,069,048 unique domain records remain: 875,958 apex domains, 2,447 parent domains associated with collected subdomains, and 190,643 subdomains. Domains matched by the scanner's provider rules are labeled CDN, and the rest are labeled non-CDN. All scans ran from a single cloud vantage point.

\textbf{Scanning procedure.} Phase~1 resolves each domain and its collected subdomains over DNS and groups them by returned address. A domain--address pair in which the address also serves another domain is a candidate for condition A1. Phase~2 opens a TLS connection to that address with the server name indication (SNI) and HTTP Host both set to a probe domain that also resolves to the address. This exercises condition A4. The scanner fetches the response and compares it with the probe domain's public baseline. A match indicates that the edge routed the connection to the probed tenant rather than returning a generic edge response, which exercises condition A3. We compute similarity with SimHash~\cite{sadowski2007simhash} at a threshold of 0.90.

\textbf{Probe sources.} Probe domains come from CDN providers, same-IP co-tenants, and a fallback for isolated addresses that reuses the provider set. The analysis contains 18,025,068 successful probe records: 14,354,459 (79.6\%) use provider domains and 3,670,609 (20.4\%) use co-tenant domains.

\textbf{Outcome classification.} HIGH requires a successful TLS handshake and HTTP content that matches the probe-domain baseline. MEDIUM requires a successful handshake and an HTTP response without a baseline match. LOW requires TLS acceptance without HTTP verification, and NONE means the handshake is rejected. We treat HIGH, MEDIUM, and LOW as exposed because all three accept an external SNI. Domain-level rates use all 1,069,048 measured domains.

\textbf{Laboratory testbeds.} We test network inspection with Zeek~\cite{bro1998system,zeek2026} and Suricata~\cite{oisf2026suricata} in IDS and inline-IPS deployments. Address-level enforcement uses OPNsense~\cite{opnsense2026} and pfSense~\cite{pfsense2026}, and the DNS-bound control is Windows ZTDNS~\cite{microsoft2025ztdns}. The inspection and address-level runs use a shared-address endpoint in which \(D_{\mathit{allowed}}\) and \(D_{\mathit{malicious}}\) resolve to the same address, with an NGINX/OpenSSL server and an OpenSSL client for ECH. The ZTDNS runs use a CDN-backed pair at one edge address and a Windows-native ECH client with \texttt{ech\_accepted=true}. These runs allow us to verify condition A2. A run succeeds when the client retrieves the content of \(D_{\mathit{malicious}}\) and, for ECH runs, the handshake is accepted.

\subsection{Routing-Plane Exposure}

\begin{figure}[t]
\centering
\includegraphics[width=0.9\columnwidth]{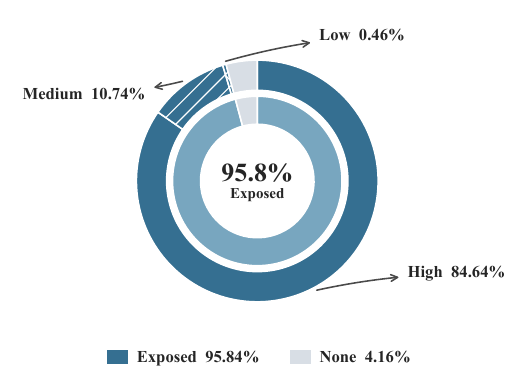}
\caption{Outcome distribution across domains.
}
\label{fig:exposure}
\end{figure}

Figure~\ref{fig:exposure} shows that 904,789 of the 1,069,048 measured domains (84.6\%) reach HIGH. At HIGH, the edge accepts the external SNI and returns content matching the probe-domain baseline. A further 114,820 domains (10.74\%) are MEDIUM and 4,931 (0.46\%) are LOW. The remaining 44,508 domains (4.16\%) reject the external SNI. In total, 1,024,540 domains (95.8\%) accept a TLS connection that carries an external SNI.

The HIGH rate provides the evidence that the edge routed the connection to the named tenant. The 95.8\% rate counts every domain for which at least one external connection identity was accepted, while HIGH additionally requires the response to match the named tenant.

Exposure reaches 99.26\% for the 506,763 CDN domains and 92.75\% for the 562,285 non-CDN domains, a difference of 6.51 percentage points. The higher CDN rate is consistent with edge nodes that perform TLS termination for multiple tenants. Non-CDN exposure exceeds 90\%, so the pattern extends to non-CDN deployments.

\finding{1}{Of 1,069,048 domains, 95.8\% accept an external SNI and 84.6\% return content matching the named tenant. Exposure is 99.26\% for CDN and 92.75\% for non-CDN domains, so the pattern extends to non-CDN hosting.}

\begin{figure}[t]
\centering
\includegraphics[width=1\columnwidth]{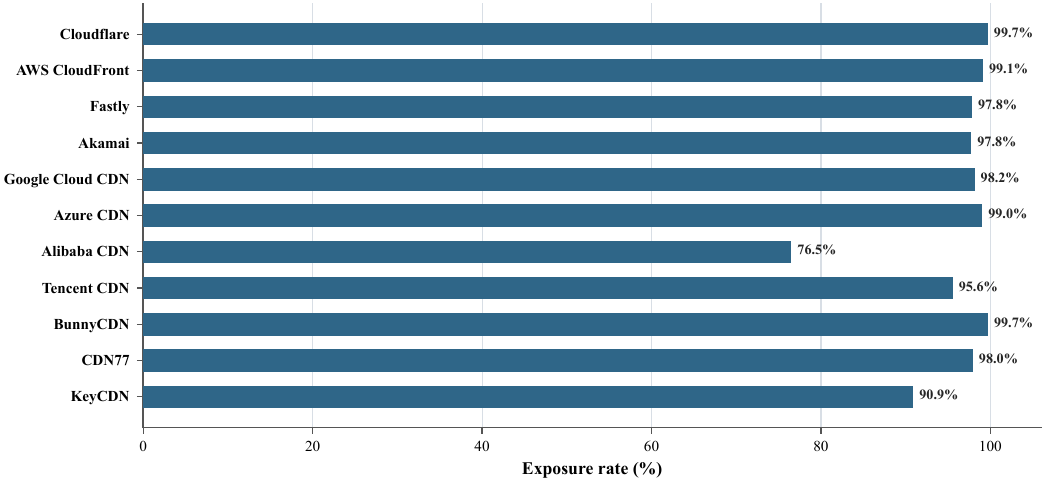}
\caption{Exposure rate by CDN provider.}
\label{fig:cdn_provider}
\end{figure}

Figure~\ref{fig:cdn_provider} compares eleven CDN providers. Ten exceed 90\% exposure. Cloudflare reaches 99.72\% across 385,145 domains, AWS CloudFront reaches 99.14\% across 49,880 domains, and Azure CDN reaches 99.05\% across 6,664 domains. Alibaba CDN is the outlier at 76.52\% across 3,629 domains. The spread shows that exposure varies by provider.

For the co-location analysis, we use the registrable domain (eTLD+1) as a proxy for tenant identity. An address is cross-tenant when it serves more than one registrable domain, and an address is exploitable when at least one probe returns another tenant's content.

\begin{figure}[t]
\centering
\includegraphics[width=0.9\columnwidth]{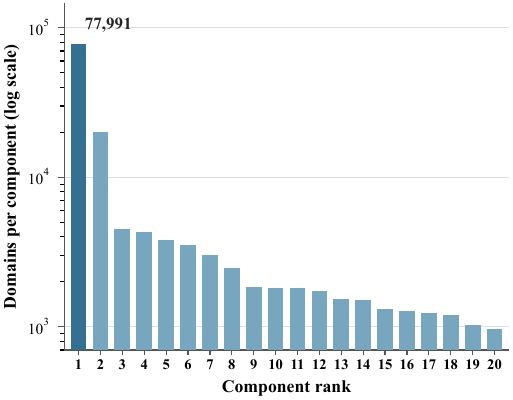}
\caption{The 20 largest connected components of the same-IP co-location graph, ordered by size. Bars show component sizes on a logarithmic scale, and the largest component contains 77,991 domains.}
\label{fig:attack_graph}
\end{figure}

The 1,024,540 exposed domains resolve to 520,340 unique addresses, and the distribution is concentrated. The busiest address hosts 16,985 exposed domains, and the twenty busiest host an average of 3,541. To characterize how far this sharing extends, we build a graph in which two domains are connected when they share an address. The graph contains 758,232 domains in 81,466 connected components. The largest component contains 77,991 domains. Of these components, 77,375 span more than one registrable domain and contain at least one exposed domain. Together they contain 666,413 exposed domains (Fig.~\ref{fig:attack_graph}).

The component structure describes the reach available through repeated one-hop steps, each requiring a permitted address. Our probes confirm the one-hop relation at individual addresses: presenting \(D_{\mathit{malicious}}\) as the SNI and Host to an address obtained through \(D_{\mathit{allowed}}\) returns the content of \(D_{\mathit{malicious}}\). With additional permitted addresses, the same step extends across a cross-tenant component.

\begin{figure}[t]
\centering
\includegraphics[width=\columnwidth]{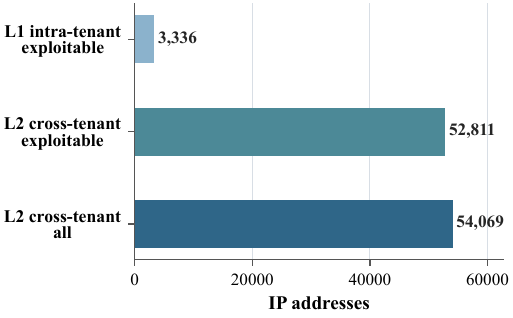}
\caption{Non-CDN addresses host multiple domains. L1 addresses serve names under one registrable domain, and L2 addresses serve multiple registrable domains.}
\label{fig:non_cdn}
\end{figure}

The same relation appears outside conventional CDN deployments. Among 58,854 non-CDN addresses that host more than one domain, 54,069 serve more than one registrable domain. The other 4,785 serve names under a single registrable domain. Of the 54,069 cross-tenant addresses, 52,811 (97.7\%) contain at least one confirmed cross-tenant path, covering 330,369 exposed domains (Fig.~\ref{fig:non_cdn}). The co-location structure extends from CDN infrastructure to non-CDN shared hosting.

\finding{2}{The same-IP graph contains 666,413 exposed domains in cross-tenant components, and 97.7\% of cross-tenant non-CDN addresses contain at least one confirmed cross-tenant path.}

Exposure is 97.90\% for apex domains, 92.15\% for subdomain parents, and 86.42\% for subdomains. GeoLite2~\cite{maxmind2026geolite2} places the measured edge addresses on six continents by registration geography, and the exposure rate over domain--address pairs ranges from 85.80\% to 99.07\%. Appendix~\ref{app:supp} reports the remaining views.

\subsection{Defense Evaluation}

We further evaluate the security risk of the two planes of Section~\ref{sec:model}. Zeek and Suricata inspect the visible connection identity on the routing path (\(P_{\mathit{dpi}}\)). OPNsense and pfSense enforce address-level policies, and ZTDNS converts a trusted DNS decision (\(P_{\mathit{dns}}\)) into an endpoint-side permit on the returned address (\(P_{\mathit{fw}}\)). We test a cleartext connection identity and ECH for the hidden tenant, two visibility conditions that change what a path observer sees, not what the shared edge knows.

We ask whether the tested defenses can bind the connection identity that selects a tenant to the name approved by DNS. Without ECH, the connection presents \(D_{\mathit{malicious}}\) in the SNI. Zeek records that name, and Suricata in the IDS deployment generates a corresponding alert. Suricata in the inline-IPS deployment can drop \(D_{\mathit{malicious}}\) while leaving \(D_{\mathit{allowed}}\) reachable, because the two names remain distinguishable at the SNI layer. Among the routing-plane tools, this is the only tested configuration in which the hidden tenant can be blocked selectively.

With ECH, the inner SNI is encrypted. Zeek and Suricata observe only the outer public name and the destination IP address, so rules that target the inner SNI no longer match \(D_{\mathit{malicious}}\). The remaining observable signals are the shared address and ECH extension 65037~\cite{iana_tls_extensiontype_values,rfc9849}. A rule on either signal cannot block the hidden tenant without also blocking traffic to \(D_{\mathit{allowed}}\). Structural parsing identifies the ECH extension, while the inner name remains encrypted. The controlled ECH deployment completed the handshake, so the inspection results reflect an ECH connection.

OPNsense and pfSense classify traffic by IP address and port. Because \(D_{\mathit{allowed}}\) and \(D_{\mathit{malicious}}\) share the same edge address, these controls cannot separate the two tenants. Blocking the address also blocks the anchor.

Windows ZTDNS blocks a direct connection to \(D_{\mathit{malicious}}\), as recorded by BlockedConnections Event~ID~2. Resolving \(D_{\mathit{allowed}}\) places the shared edge address in the permitted address set, which supplies condition A2. A later ordinary-TLS connection to that address presents \(D_{\mathit{malicious}}\) in the SNI and retrieves its content. The permit derived from \(D_{\mathit{allowed}}\) is therefore reused by another tenant on the same address.

The same reuse occurs with ECH. After obtaining a valid ECHConfig for \(D_{\mathit{malicious}}\) and resolving \(D_{\mathit{allowed}}\), a Windows-native ECH client completes the connection with \texttt{ech\_accepted=true} and retrieves the content of \(D_{\mathit{malicious}}\). ECH is not required for the bypass, because ordinary TLS already reuses the shared-address permit. Its role is to hide the connection identity from network inspection. Once \(D_{\mathit{allowed}}\) permits the shared address, the external-DoH and ECH path reaches \(D_{\mathit{malicious}}\) while concealing both the DNS query content and the inner SNI. DoH could strengthen concealment~\cite{rfc8484}. The bypass itself comes from the reused address permit.

\begin{table}[t]
\centering
\caption{Laboratory outcome by defense and connection identity.
A is \(D_{\mathit{allowed}}\) and B is \(D_{\mathit{malicious}}\).
Both names resolve to the same edge address in each testbed.}
\label{tab:defense}
\small
\setlength{\tabcolsep}{4pt}
\renewcommand{\arraystretch}{1.2}
\begin{tabularx}{\columnwidth}{
    @{}
    >{\raggedright\arraybackslash}p{0.2\columnwidth}
    >{\raggedright\arraybackslash}p{0.19\columnwidth}
    >{\raggedright\arraybackslash}X
    >{\raggedright\arraybackslash}X
    @{}
}
\toprule
Plane & Defense & Cleartext SNI & ECH \\
\midrule
\multirow{3}{=}{\raggedright\textit{Routing inspection}}
& Zeek
& Logs B's SNI
& Logs outer name only \\
& Suricata (IDS)
& Alerts on B's SNI
& No inner SNI alert \\
& Suricata (IPS)
& Drops B (A intact)
& SNI rule misses; IP/65037 rule hits both \\
\midrule
\multirow{3}{=}{\raggedright\textit{Authorization enforcement}}
& OPNsense
& IP-only, non-selective
& IP-only, non-selective \\
& pfSense
& IP-only, non-selective
& IP-only, non-selective \\
& ZTDNS
& Direct B blocked; reused after A
& Reused after A (confirmed) \\
\bottomrule
\end{tabularx}
\end{table}

Table~\ref{tab:defense} summarizes the outcomes. No tested control performs the cross-plane check \(D_{\mathit{conn}}\in N(D_{\mathit{req}})\). The inspection tools observe the connection identity without the DNS-approved name, and the enforcement tools act on the address without the tenant identity. Under ECH, the path observer loses the connection identity. Cleartext inline IPS can block the hidden tenant at the SNI layer, but that check is unavailable under ECH. The ZTDNS address permit remains reusable in both visibility conditions. A domain-level DNS decision becomes an address-level permit, and the permit does not preserve the domain binding when a later connection selects another tenant behind the same address.

\finding{3}{No tested control binds the connection identity to the DNS-approved name. Cleartext inline IPS can block the hidden tenant, but the check is unavailable under ECH, and the ZTDNS address permit remains reusable.}

\section{Threats to Validity}\label{val}

\textbf{Exposure-classification validity.}
The 95.8\% exposure rate counts a domain once at least one probe accepts a TLS connection carrying an external SNI. It does not depend on the SimHash threshold, which only separates HIGH from MEDIUM. HIGH adds a content match and covers 84.6\% of domains. MEDIUM and LOW establish TLS without a confirmed tenant match, are reported separately, and are never promoted to HIGH. Domain-level rates use all 1,069,048 measured domains, including the 44,508 that reject the external SNI.

\textbf{Measurement-scope validity.}
The measurement scan contains no protective DNS or firewall, so its rates measure the exposure that DDA requires rather than the bypass rate of a deployed control. All scans ran from a single cloud vantage point, and edge behavior can depend on the client's network location. The rates describe the Tranco-derived domain set and its collected subdomains, not the full domain space. Provider-domain probes supply 79.6\% of successful records and co-tenant probes 20.4\%, which reduces reliance on one probe category but does not add a second vantage point. The provider comparison reports observed differences without isolating the routing or termination policy behind them. GeoLite2 labels describe registration geography, not the path of a connection.

\textbf{Shared-tenancy and reach validity.}
We use the registrable domain (eTLD+1) as a proxy for tenant identity. A cross-tenant address serves more than one registrable domain, so the label can count names from one organization as separate tenants and can miss multiple tenants served under one registrable domain. The CDN/non-CDN split follows the scanner's provider rules rather than a provider-independent classification. The co-location graph connects domains that share an address, and component size describes connected exposure rather than single-permit reach. Each one-hop step requires an address permitted through an allowed anchor, and our probes confirm the one-hop relation at individual addresses, not every multi-hop path in a component. The 666,413 figure counts exposed domains in cross-tenant components, and the 97.7\% non-CDN figure counts cross-tenant addresses with at least one confirmed path.

\textbf{Attack-deployment validity.}
A real deployment requires an approved anchor whose address also serves the hidden tenant (A1). The measurement quantifies the reachability that such a co-located destination would have, and acquiring the hidden tenant is ordinary onboarding on the shared provider. The Internet scan exercises A1, A3, and A4. A2 requires a DNS-bound control where resolving the anchor places the shared address in the permitted set. We do not evaluate session-state reuse separately, because the construction does not depend on stateful inspection. Direct-IP access lies outside the trusted-DNS scope, and we treat it as a boundary rather than a variant of DDA.

\textbf{Defensive-evaluation validity.}
The defense experiments use controlled testbeds rather than a production enterprise deployment. Zeek, Suricata, OPNsense, and pfSense run against a shared-address endpoint, and the ZTDNS runs use a CDN-backed pair. The stack is representative but not exhaustive, and ZTDNS is the only DNS-bound control we test. Its address-level permit is reused on a shared edge, so the same reuse is expected for other controls that admit resolver-returned addresses without binding them to the connection identity, although other implementations may add checks we did not exercise. The ECH runs cover an NGINX/OpenSSL server with an OpenSSL client in the shared-address testbed and a Windows-native client against the CDN-backed pair. Structural parsing identifies the ECH extension in these runs, not the inner name, so extension-based detection supports blocking ECH traffic as a class rather than selecting the hidden tenant.

\section{Related Work}

Domain fronting exposes a high-reputation domain in DNS and TLS SNI while directing the encrypted HTTP Host header to another destination, provider-side SNI–Host consistency checks were designed to restrict it, yet a measurement of 30 CDNs found 22 still permitted some form of it~\cite{fifield2015blocking,subramani2024discovering}. Domain borrowing instead presents another domain consistently as both SNI and Host, relying on gaps in domain ownership verification~\cite{ding2021domain}, domain shadowing relies on CDN-internal front-to-backend mapping~\cite{wei2021domain}, and domain hiding used early ESNI with a visible outer name, a historical behavior not equivalent to standardized ECH~\cite{hunstad2020domainhiding}. IP-level blocking of shared edges also causes collateral impact on benign domains~\cite{Zolfaghari2016Practical}. These techniques place the domain difference within the connection or depend on provider-specific weaknesses, indicating that the domain observed by DNS, the connection identity, and the shared edge address are not a fixed one-to-one correspondence.

Shared addressing is a basic deployment model for CDNs and virtual hosting, in which one address hosts many domains and TLS SNI with the HTTP authority select the service~\cite{rfc7230}, measurements such as Satellite have characterized the resulting many-to-many relationship between domains, addresses, and services~\cite{scott2016satellite,hoang2020web}. Protective DNS enforces policy at resolution time, and Windows ZTDNS blocks outbound IP traffic by default and admits addresses returned by a trusted DNS service~\cite{microsoft2025ztdns}: the object actually enforced is the IP, so when that address belongs to a shared edge, the domain-level decision and the address-level permission operate at different granularities. Prior measurement work has paid limited attention to whether this DNS-derived permission remains bound to the originally approved domain.

DoH hides query names from on-path observers~\cite{Hoffman2018DoH} and ECH encrypts the server name in the inner ClientHello~\cite{Rescorla2026ECH}, but neither hides the connection itself, and traffic features may still reveal the target service~\cite{Trevisan2023Attacking,hoang2021domain}, on a shared IP an address rule applies to many domains simultaneously, and detecting ECH use does not reveal the encrypted inner domain. The joint effect of encrypted domain information, shared edges, and DNS-derived connection permission has not been directly examined.

\section{Conclusion}

We presented the Domain Decoupling Attack (DDA), which exploits the validation gap between the domain approved during DNS resolution and the identity that selects a tenant at connection time. An allowed anchor places a shared edge address in the permitted set, and a later connection to that address selects a different tenant while keeping the connection identity and HTTP Host consistent. No online check binds the tenant-selecting identity to the name that justified the permit. A measurement of 1,069,048 domains produced 18,025,068 successful probes: 95.8\% accepted an external SNI, 84.6\% returned content matching the named tenant, and exposure was 99.26\% for the 506,763 CDN domains and 92.75\% for the 562,285 non-CDN domains. The co-location graph contains 666,413 exposed domains in cross-tenant components. No tested defense performs the cross-plane binding check: cleartext inline IPS can block the hidden tenant only at the SNI layer, ECH hides that signal, and the ZTDNS address permit remains reusable in both cases. The gap conceals command-and-control and data-exfiltration traffic and supports censorship circumvention.

\section*{Ethical Considerations}

Our measurements use standard TLS handshakes and a small number of HTTP requests to publicly accessible edge services, without authentication, payload delivery, or disruptive traffic. The study examines intended multi-tenant routing behavior and uses no production traffic. IP addresses and domains are anonymized in figures, and no personal data is collected. As the issue arises from shared-edge architecture rather than a single vendor-specific flaw, disclosure is unusable. We focus on measuring exposure, demonstrating the attack, and evaluating potential mitigations.

\bibliographystyle{ACM-Reference-Format}
\bibliography{reference}

\appendix
\section{Supplementary Measurement Results}\label{app:supp}

\begin{figure}[ht]
\centering
\includegraphics[width=0.8\columnwidth]{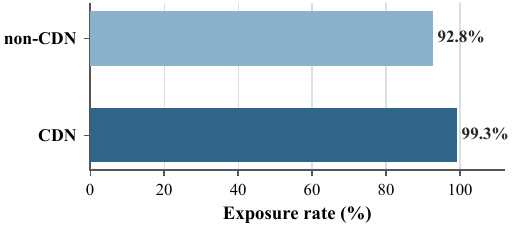}
\caption{Exposure for CDN and non-CDN domains. CDN domains (n=506,763) reach 99.26\% exposure and non-CDN domains (n=562,285) reach 92.75\%. Bars give the exposed share.}
\label{fig:app_exposure_type}
\end{figure}

\textbf{CDN and non-CDN domains.} Figure~\ref{fig:app_exposure_type} compares the exposure result for CDN and non-CDN domains. Of the 562,285 non-CDN domains, 521,521 (92.75\%) accept an external SNI, and 503,019 of the 506,763 CDN domains (99.26\%) do the same. The gap widens at the HIGH level: 437,278 non-CDN domains (77.8\%) and 467,511 CDN domains (92.3\%) return content that matches the probe-domain baseline. Both groups are instances of the same shared-edge pattern.

\begin{figure}[ht]
\centering
\includegraphics[width=\columnwidth]{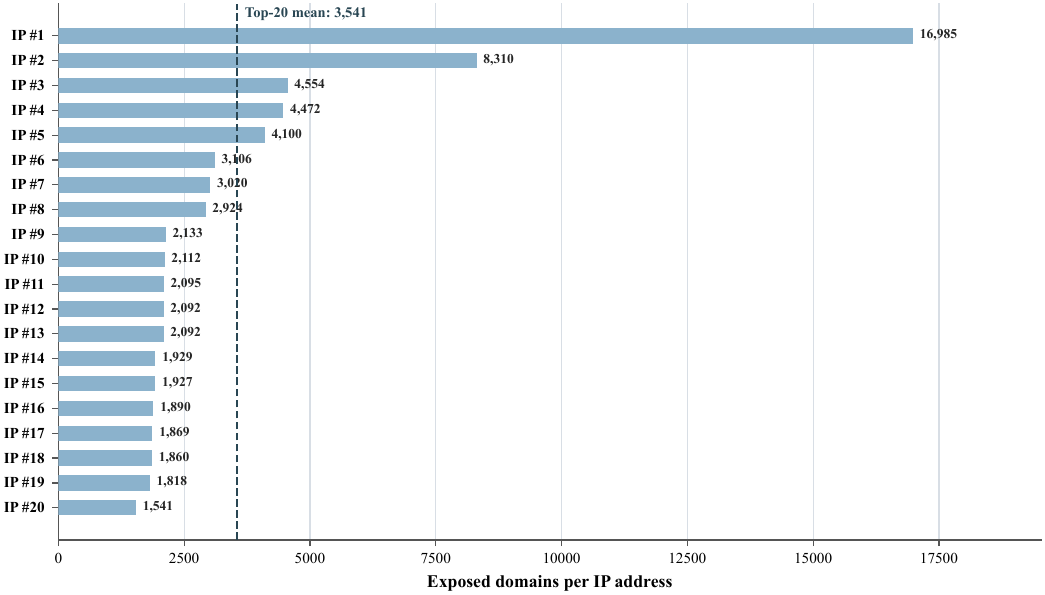}
\caption{Exposed domains at the 20 busiest edge addresses. Addresses are anonymized as IP \#1 to IP \#20 and ordered by exposed-domain count. The dashed line marks the top-20 mean of 3,541 exposed domains.}
\label{fig:app_ip_hotspots}
\end{figure}

\textbf{Address concentration.} Figure~\ref{fig:app_ip_hotspots} ranks the 20 addresses with the most exposed domains. The 1,024,540 exposed domains resolve to 520,340 unique addresses, and the distribution is concentrated. The busiest address hosts 16,985 exposed domains, and the top 20 host 3,541 on average. Fifteen of the 20 expose every domain observed at the address, and the lowest exposure rate in the group is 96.6\%. The figure plots exposed domains per address rather than probe volume, and anonymizes addresses as IP \#1 to IP \#20.

\begin{figure}[ht]
\centering
\includegraphics[width=0.85\columnwidth]{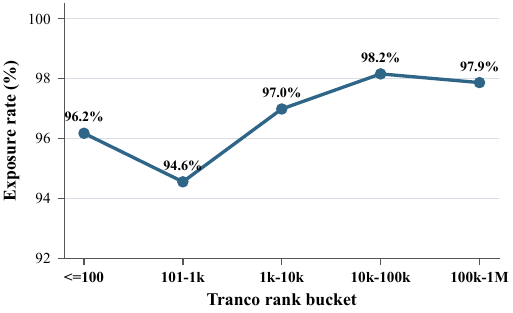}
\caption{Overall exposure by Tranco rank bucket. Buckets cover 878,405 apex and parent domains with a Tranco rank. Subdomains are excluded because they have no independent rank. The line gives the exposure rate across buckets.}
\label{fig:app_popularity}
\end{figure}

\textbf{Popularity.} Figure~\ref{fig:app_popularity} reports the overall exposure rate across Tranco rank buckets. The buckets cover 878,405 apex and parent domains with a Tranco rank and exclude the 190,643 subdomains, which have no independent rank. The rate is 96.18\% in the top 100, 94.56\% in 101--1k, 96.99\% in 1k--10k, 98.16\% in 10k--100k, and 97.87\% in 100k--1M. The spread is 3.6 percentage points and shows no monotonic trend. The top-100 bucket contains 157 domains, and the 100k--1M bucket contains 791,525.

\begin{figure}[ht]
\centering
\includegraphics[width=0.85\columnwidth]{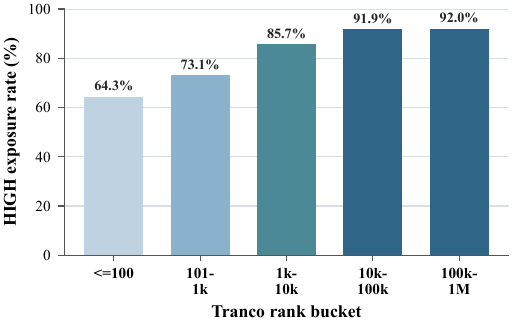}
\caption{HIGH-level rate by Tranco rank bucket. Bars give the share of domains in each bucket classified HIGH.}
\label{fig:app_high_rate}
\end{figure}

\textbf{HIGH share.} Figure~\ref{fig:app_high_rate} shows the HIGH-level rate within each rank bucket. HIGH accounts for 101 of 157 domains in the top 100 (64.3\%), 1,007 of 1,378 in 101--1k (73.1\%), 7,775 of 9,074 in 1k--10k (85.7\%), 70,087 of 76,271 in 10k--100k (91.9\%), and 728,506 of 791,525 in 100k--1M (92.0\%). The HIGH share rises from the most popular bucket to the least popular one and reaches its maximum in the 100k--1M bucket. Overall exposure is nearly flat across buckets (Figure~\ref{fig:app_popularity}), so the gap between exposure and HIGH is largest in the most popular buckets.

\begin{figure}[ht]
\centering
\includegraphics[width=\columnwidth]{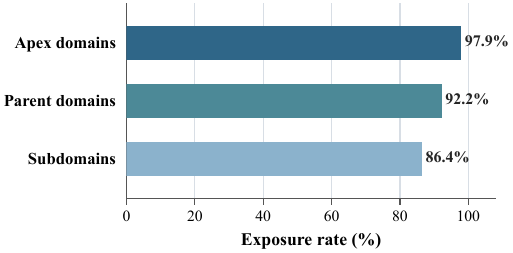}
\caption{Exposure by DNS depth. Apex domains (n=875,958), subdomain parents (n=2,447), and subdomains (n=190,643). Bars give the exposed share.}
\label{fig:app_dns_depth}
\end{figure}

\textbf{DNS depth.} Figure~\ref{fig:app_dns_depth} compares apex domains, subdomain parents, and subdomains. Apex domains reach 97.90\% exposure (857,535 of 875,958), subdomain parents reach 92.15\% (2,255 of 2,447), and subdomains reach 86.42\% (164,750 of 190,643). The rate decreases with DNS depth.

\begin{figure}[ht]
\centering
\includegraphics[width=0.9\columnwidth]{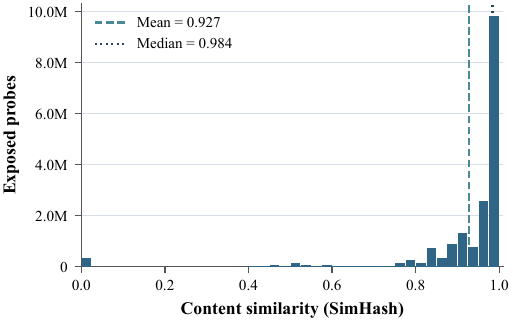}
\caption{SimHash similarity for probes that accepted an external SNI. The histogram covers 18,025,068 successful probe records. The classification threshold is 0.90, and the median is 0.98.}
\label{fig:app_similarity}
\end{figure}

\textbf{Similarity.} Figure~\ref{fig:app_similarity} shows the distribution of SimHash similarity for the 18,025,068 successful probe records, all of which accepted an external SNI. The distribution is concentrated near 1.0, with a mean of 0.93, a median of 0.98, a 25th percentile of 0.92, and a 75th percentile of 1.00. The 0.90 threshold used for the HIGH classification lies below the 25th percentile.

\begin{figure}[ht]
\centering
\includegraphics[width=0.9\columnwidth]{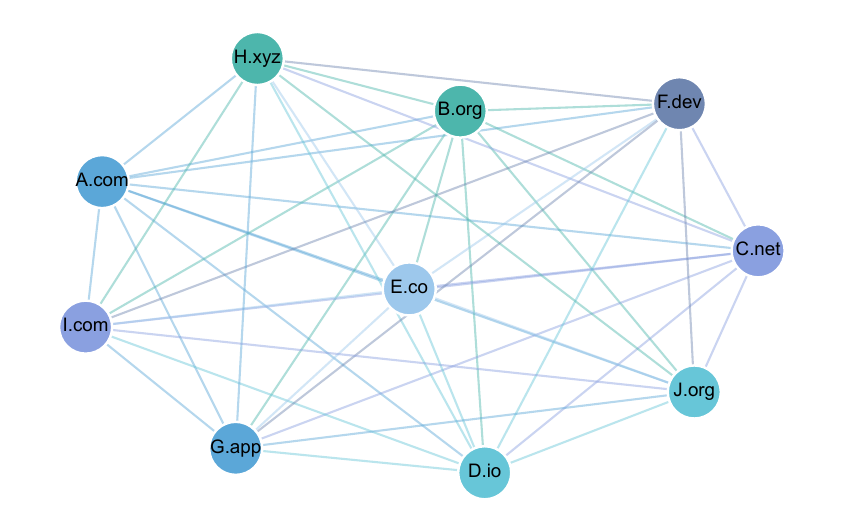}
\caption{An anonymized 10-node subset of a cross-tenant component. Labels are placeholders, and an edge denotes a shared address. All nodes in this subset are exposed.}
\label{fig:app_cluster}
\end{figure}

\textbf{Co-location component.} Figure~\ref{fig:app_cluster} shows a 10-node anonymized subset of one cross-tenant component in the same-IP graph. Each edge joins two domains that share at least one address, and every node in the subset is exposed. The full graph contains 81,466 connected components and 77,375 cross-tenant components with at least one exposed domain. The figure shows how shared addresses connect otherwise unrelated tenants. The component structure describes reach through repeated one-hop steps, each requiring a permitted address.

\begin{figure}[ht]
\centering
\includegraphics[width=0.9\columnwidth]{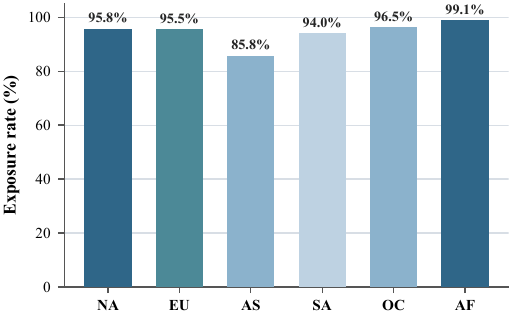}
\caption{Exposure over domain--address pairs by registration continent. Addresses with unknown geography are excluded. Bars give the exposed share of pairs in each continent.}
\label{fig:app_continent}
\end{figure}

\begin{figure}[ht]
\centering
\includegraphics[width=\columnwidth]{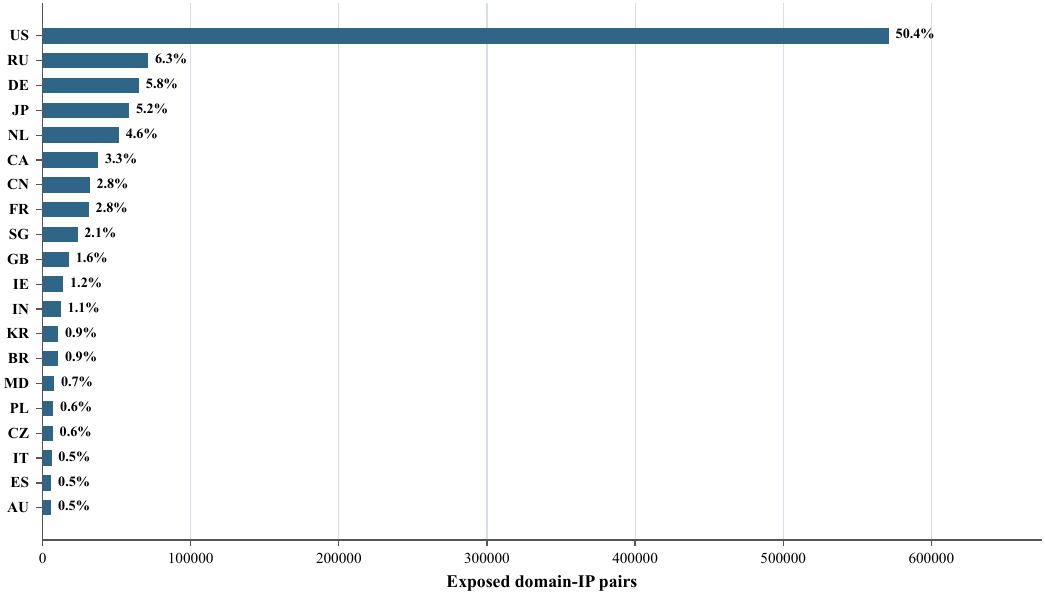}
\caption{Top 20 countries by exposed domain--address pairs. Unknown geography is excluded. Labels give each country's share of the 1,132,784 geo-located exposed pairs.}
\label{fig:app_country}
\end{figure}

\textbf{Location.} Figure~\ref{fig:app_continent} reports exposure over domain--address pairs by the registration continent of the edge address, using GeoLite2 labels~\cite{maxmind2026geolite2}. Addresses with unknown geography are excluded. The rate is 95.77\% in North America, 95.50\% in Europe, 85.80\% in Asia, 93.98\% in South America, 96.48\% in Oceania, and 99.07\% in Africa. Exposure exceeds 85\% on every continent. A domain can contribute to more than one continent when it resolves to addresses in different regions. Figure~\ref{fig:app_country} ranks the top 20 countries by exposed domain--address pairs after excluding unknown geography. The United States accounts for 50.4\% of the 1,132,784 geo-located exposed pairs, followed by Russia (6.3\%), Germany (5.8\%), and Japan (5.2\%). The ranking follows the registration geography of edge addresses rather than the location of users or domain owners.

\FloatBarrier

\end{document}